\documentclass[doublecol]{epl2} 

\usepackage[utf8x]{inputenc}
\usepackage{blindtext}
\usepackage{amstext}
\usepackage{amsmath}
\usepackage{amssymb}
\usepackage{amsfonts}
\usepackage{soul}

\renewcommand{\vec}[1]{\boldsymbol{#1}}  
\newcommand*{\diff}{\mathrm{d}}
\newcommand{\del}{\partial}
\newcommand{\Lap}{\Delta}
\let\sphi\phi 
\let\phi\varphi 

\definecolor{Red}{rgb}{0.9,0.0,0.1}
\definecolor{Blue}{rgb}{0.1,0.0,0.9}
\definecolor{Darkblue}{rgb}{0.22,0.33,0.64}

\title{Secondary polygonal instability of buckled spherical shells}

\author{Sebastian Knoche\inst{1} \and Jan Kierfeld\inst{1}}
\shortauthor{S. Knoche and J. Kierfeld}

\institute{                    
  \inst{1} Department of Physics, Technische Universit\"{a}t Dortmund, 44221 Dortmund, Germany
}


\abstract{ 
  When a spherical elastic capsule is deflated, it first buckles axisymmetrically and subsequently loses its axisymmetry in a secondary instability, where the dimple acquires a polygonal shape. We explain this secondary polygonal buckling in terms of wrinkles developing at the inner side of the dimple edge in response to compressive hoop stress. Analyzing the axisymmetric buckled shape, we find a compressive hoop stress with parabolic stress profile at the dimple edge. We further show that there exists a critical value for this hoop stress, where it becomes favorable for the membrane to buckle out of its axisymmetric shape, thus releasing the compression. The instability mechanism is analogous to the formation of wrinkles under compressive stress. A simplified stability analysis allows us to quantify the critical stress for secondary buckling. Applying this secondary buckling criterion to the axisymmetric shapes, we can determine the critical volume for secondary buckling. Our analytical result is in close agreement with existing numerical data.
}

\begin{document}

\maketitle

\section{Introduction}

\begin{figure}[t]
  \centerline{\includegraphics[width=85mm]{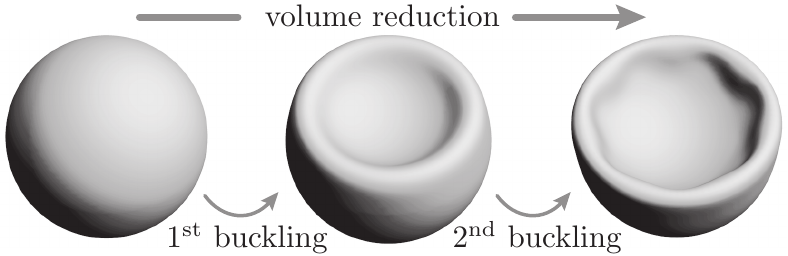}}
  \caption{During deflation an elastic  capsule first buckles into an axisymmetric shape and then undergoes a secondary buckling where the dimple acquires a polygonal shape.}
\label{fig:3Dplots}
\end{figure}

All spherical elastic shells, such as sports and toy balls or microcapsules, exhibit a qualitatively identical deformation behaviour upon deflation: At small deflation, the capsule remains spherical. Below a critical volume, the classical buckling instability occurs, and an axisymmetric dimple forms \cite{Timoshenko1961, Landau1986, Ventsel2001}. Finally, this dimple loses its axisymmetry in a secondary instability, resulting in a polygonal buckled shape (see fig.\ \ref{fig:3Dplots}). This deformation behaviour is seen in daily life on a macroscopic scale for elastomer balls, on the microscale in experiments on microcapsules \cite{Quilliet2008, Datta2010, Datta2012}, as well as in computer simulations based on triangulated surfaces \cite{Quilliet2008, Quilliet2010, Vliegenthart2011, Quilliet2012} or finite element methods \cite{Vella2011,Vaziri2008,Vaziri2009}. The same sequence of an axisymmetric buckling instability followed by a secondary polygonal buckling instability also occurs when a dimple is formed by indenting the capsule with a point force \cite{Pauchard1998, Vaziri2008, Vaziri2009}, when the capsule is pressed between rigid plates \cite{Vaziri2009} or when the capsule adheres to a substrate \cite{Komura2005}.

The first buckling transition, where an axisymmetric dimple forms, is well understood. Linear shell theory can be successfully used to calculate the onset of instability of the spherical shape \cite{Landau1986,Ventsel2001}. Furthermore, nonlinear shell theory has been used to investigate the post-buckling behaviour, which revealed that the buckled shape is unstable with respect to further volume reduction if the pressure is controlled \cite{Landau1986,Koiter1969,Knoche2011}. Numerical analyses of nonlinear shape equations have been used to characterise the bifurcation behaviour of axisymmetric shapes \cite{Bauer1970,Knoche2011}. The results of ref.\ \cite{Knoche2011} show that the first buckling transition exhibits a bifurcation behaviour that is analogous to a first order phase transition, with the volume for the onset of instability (the spinodal) differing from the critical volume where the elastic energy branches of spherical and buckles shapes cross.

In contrast, the theory of the secondary buckling transition, where the dimple loses its axisymmetry, has remained mostly phenomenological based on the existing results from computer experiments \cite{Quilliet2008, Quilliet2010, Quilliet2012}. A theory rationalizing the underlying mechanism and predicting the critical volume of secondary buckling is still lacking. In this Letter, we show that secondary buckling  is caused by compressive hoop stresses that occur in the inner neighbourhood of the dimple edge after  axisymmetric buckling. In order to release the compressive stress, the circumferential fibres buckle out of their circular shape if the hoop stress reaches a critical value; this instability is similar to wrinkling under compressive stress \cite{Timoshenko1961} or comparable to the Euler buckling of straight rods \cite{Landau1986}. The quantitative investigation of the secondary buckling transition therefore consists of two steps: Firstly, determining the stress distribution in the axisymmetric buckled configuration, and secondly, finding the critical compressive stress at which the axisymmetric configuration loses its stability.

The first task can be accomplished by numerical integration of the shape equations derived from nonlinear shell theory \cite{Knoche2011} or by an analytic approach based on the ideas of ref.\ \cite{Pogorelov1988}. The second task necessitates an analysis of the stability equations of shallow shells \cite{Ventsel2001}. In this Letter, we focus  on the mechanism of  secondary buckling and  the resulting parameter dependencies of the critical buckling volume. Detailed calculations  and numerical work will be published elsewhere \cite{Knoche2014}.

\section{Axisymmetric buckling of capsules}


\begin{figure}[t]
  \centerline{\includegraphics[width=85mm]{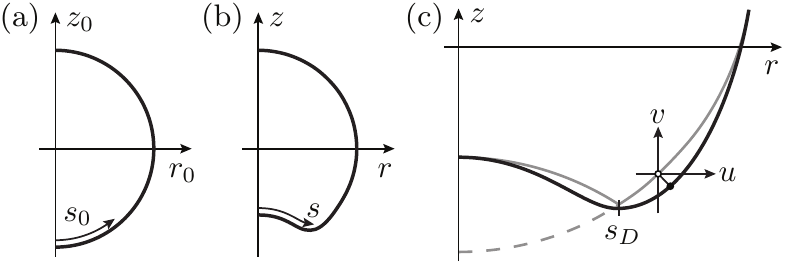}}
  \caption{Geometry of the axisymmetric midsurface. (a) Undeformed shape (always with index ``0''), (b) deformed shape, (c) Pogorelov model: Close-up of the smooth final shape (black) with small displacements $(u, v)$ to the isometric shape (grey).}
\label{fig:midsurface}
\end{figure}

All axisymmetric capsule shapes can be obtained as solutions of shape equations derived from nonlinear shell theory \cite{Knoche2011}.  Figure  \ref{fig:midsurface} shows the parametrisation of the capsule midsurface. The spherical reference configuration is  given by $r_0(s_0) = R_0 \sin(s_0/R_0)$ and $z_0(s_0) = - R_0 \cos(s_0/R_0)$ in arclength parametrisation. Curvatures in meridional and circumferential direction are equal: $\kappa_{s_0} = \kappa_{\phi_0} = 1/R_0$.

Upon axisymmetric deformation, the midsurface undergoes stretching and bending. We measure the stretches in meridional and circumferential direction by $\lambda_s = \diff s / \diff s_0$ and $\lambda_\phi = r/r_0$, respectively, with corresponding strains  $e_s = \lambda_s - 1$ and $e_\phi = \lambda_\phi -1 $. The bending strains are defined as $K_s= \lambda_s \, \kappa_s - \kappa_{s_0}$ and $K_\phi = \lambda_\phi \, \kappa_\phi - \kappa_{\phi_0}$. The deformation results in an elastic energy which is stored in the membrane. We assume the elastic energy density (measured per undeformed surface area) to be of the simple Hookean form \cite{Knoche2011}
\begin{multline}
w_S =  \frac{1}{2} \frac{EH_0}{1-\nu^2} 
  \left( e_s^2 + 2\, \nu \,e_s\, e_\phi + e_\phi^2 \right)\\
   + \frac{1}{2} E_B  \left(  K_s^2 + 2 \,\nu\, K_s \,K_\phi + K_\phi^2
   \right).
\label{elastic_energy}
\end{multline}
In this expression, $E$ is the (three-dimensional) Young modulus, $H_0$ the membrane thickness, $\nu$ the (three-dimensional) Poisson ratio which confined to $-1 \leq \nu \leq 1/2$, and $E_B = EH_0^3/12(1-\nu^2)$ the bending stiffness.

From this energy density, the meridional tension and bending moments can be derived as
\begin{align}
 \tau_s &= \frac{1}{\lambda_\phi} \frac{\del w_S}{\del e_s}
   = \frac{E H_0}{1-\nu^2} \, \frac{1}{\lambda_\phi} 
  \big( e_s + \nu\, e_\phi \big), \\
    m_s &= \frac{1}{\lambda_\phi} \frac{\del w_S}{\del K_s}
   = E_B \, \frac{1}{\lambda_\phi}\big( K_s + \nu\, K_\phi \big).
  \label{stress-strain_2D}
\end{align}
The corresponding relations for the circumferential tension and bending moment are obtained by interchanging all indices $s$ and $\phi$ in these equations. The shape is determined by the equations of force and torque equilibrium:
\begin{align}
 0 &= - \frac{\cos \psi}{r}\, \tau_\phi 
  + \frac{1}{r}\, \frac{\diff(r\,\tau_s)}{\diff s} 
   - \kappa_s\,q, \label{eq:equil1}\\
 0 &= -p + \kappa_\phi \, \tau_\phi 
   + \kappa_s \, \tau_s 
   + \frac{1}{r}\, \frac{\diff(r\,q)}{\diff s}, \label{eq:equil2}\\
 0 &=  \frac{\cos \psi}{r} \, m_\phi 
  - \frac{1}{r} \frac{\diff(r\,m_s)}{\diff s} - q.\label{eq:equil3}
\end{align}
In these equations, $q$ is the transversal shear force, and $p$ the applied normal pressure, which can also be interpreted as a Lagrange multiplier to control the capsule volume. Together with geometrical relations, these nonlinear differential equations -- called \emph{shape equations} -- can be solved numerically \cite{Knoche2011}.

For the analysis, it is convenient to introduce dimensionless quantities by using $E H_0$ as the unit for tensions and $R_0$ as the unit length. Specifically, this results in a dimensionless bending stiffness $\tilde E_B \equiv E_B/E H_0 R_0^2 = H_0^2/R_0^2 12 (1-\nu^2)$, which is the inverse of the F\"oppl-von-K\'arm\'an-number $\gamma_\text{FvK} = 1/\tilde E_B$.

For a qualitative understanding of  the shape and stress distribution of an axisymmetric buckled capsule we start with vanishing bending stiffness $E_B = 0$. Then, the equilibrium shape consists of a mirror inverted spherical cap (see Fig.\ \ref{fig:midsurface} (c), gray lines), which  is isometric to the initial spherical shape and, therefore, unstrained. For $E_B>0$,  the sharp edges of the inverted cap   give rise to an infinitely large bending energy. Hence, these sharp edges must be smoothed out.

\begin{figure}[t]
  \centerline{\includegraphics[width=86mm]{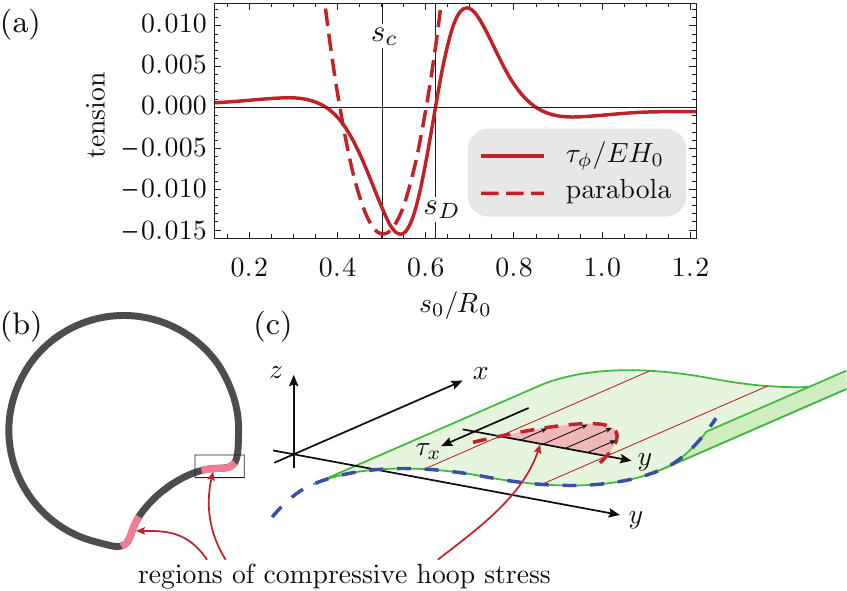}}
  \caption{(Colour online) Stress distribution (a) and shape (b) for a
    numerical solution of the shape equations at $\tilde E_B = 10^{-5}$ and
    $\Delta V / V_0 = 0.048$. The hoop tension $\tau_\phi$ is compressive
    (negative) in a narrow strip along the inner side of the dimple edge where
    we can expect wrinkles to occur. (c) Simplification of the compressed strip
    as a plate.}
\label{fig:compressive_tension}
\end{figure}

Upon  smoothing, see Fig.\ \ref{fig:midsurface} (c), the inner neighborhood of the edge is shifted to the inside, towards the axis of symmetry, and the outer neighborhood is shifted to the outside. Circumferential material fibers will  be compressed in the inner neighborhood and stretched in the outer neighborhood; but far away from the dimple edge, we expect the deformation to decay. This draws a qualitative picture of the circumferential stress distribution $\tau_\phi(s_0)$ along the arc length: It has a zero at $s_D$ (the arc length position of the edge, see fig.\ \ref{fig:midsurface}), a positive maximum for $s_0>s_D$  and a negative minimum for $s_0<s_D$; it  approaches zero for $s_0 \rightarrow 0$ and $s_0 \rightarrow \infty$ (cf.\ fig.\ \ref{fig:compressive_tension}).

Along these lines, Pogorelov constructed an analytic model for axisymmetric buckled shapes \cite{Pogorelov1988}. To describe this deformation from the isometric shape to the final smooth shape, he introduced a displacement $(u(s_0),\, v(s_0))$ (in $r$- and $z$-direction, respectively, see fig.\ \ref{fig:midsurface} (c). Assuming $u$ and $v$ to be small, linear shell theory can be employed to calculate the bending and stretching energies in the final shape by means of calculus of variations with respect to $\delta u$ and $\delta v$ (with some simplifications). From the approximate solutions $u(s_0)$ and $v(s_0)$ presented in \cite{Pogorelov1988}, analytical expressions for curvatures, tensions and stresses  of the final shape can be deduced, which are generally in good agreement with numerical solutions of our shape equations. The total elastic energy is found to be
\begin{equation}
 U_\text{Pog} = 8\pi J \frac{2^{1/4}}{3^{3/4}}  EH_0 R_0^2 
  \left( \tilde E_B \frac{\Delta V}{V_0} \right)^{3/4} (1-\nu^2)^{-1/4}
 \label{eq:Upog}
\end{equation} 
where $V_0$ is the volume of the spherical (initial) shape, $\Delta V$ the volume difference between the buckled shape and $V_0$, and $J\approx 1.15092$ a numerical constant. 

Figure \ref{fig:compressive_tension} presents  the circumferential stress distribution $\tau_\phi(s_0)$  for an axisymmetric buckled shape from the numerical solution of the shape equations; results from the Pogorelov model are in good agreement \cite{Knoche2014}. The course of $\tau_\phi$ confirms our above qualitative prediction and shows a negative peak in a narrow strip along the inner side of the dimple edge. The narrow strip is the region where wrinkles will form in order to release the compressive stress \cite{Cerda2003, Wong2006}, and it coincides with the location of wrinkles observed in simulations \cite{Quilliet2008, Quilliet2010, Vliegenthart2011, Quilliet2012}.

For the simplified stability analysis presented in the next section, the shape and stress distribution are reduced to the key features, see fig.\ \ref{fig:compressive_tension}, dashed lines. The slightly curved midsurface is approximated by a cubic parabola, which is fitted to the point where the exact midsurface has vanishing curvature $\kappa_s(s_c)=0$ (see fig.\ \ref{fig:compressive_tension}). In the vicinity of this point, the real midsurface shows a linear increase in curvature, $\kappa_s(s_0) \approx a_c (s_0 - s_c)$. The cubic parabola shall have the same slope $a_c$ of curvature. The negative peak in the hoop stress can be approximated by a parabola. It is chosen to have the same minimum value $-\tau_0$ and the same integral $\int \tau_\phi \, \diff s_0$ over the compressive part (between its roots) as the exact numerical function $\tau_\phi(s_0)$. Its centre is shifted to the point $s_c$ of vanishing meridional curvature, which is close to the minimum of the exact numerical $\tau_\phi$ function (see fig.\ \ref{fig:compressive_tension} (a)). In the following, we will neglect the meridional tension $\tau_s$, since it is small compared to $\tau_\phi$, and the curvature $\kappa_\phi$.

\section{Secondary buckling as wrinkling under locally 
compressive stress}

On the basis of these results, we will now consider the stability of a weakly curved rectangular plate (the $x$-direction corresponding to the $\phi$-direction, the $y$-direction corresponding to the $s$-direction; the plate is curved in $y$-direction). The plate is subject to a localised compression in form of a parabolic stress profile $\tau_x = -\tau_0 (1- a_p y^2)$ in a strip along the inner side of the dimple edge, see fig.\ \ref{fig:compressive_tension} (c).

Before presenting a more detailed stability analysis, we start with a scaling argument. Here we neglect curvature effects completely and approximate the compressed region by a rectangular strip of width $\Delta y \sim 1/\sqrt{ a_p}$ (identical to the compressed region) under a homogeneous compressive stress $\tau_x \sim -\tau_0$. For clamped long edges, the wrinkling wave length is given by the width, $\lambda \sim \Delta y$ \cite{Timoshenko1961}, and the resulting critical Euler buckling stress is $\tau_0 = \tau_c \sim E_B/\lambda^2 \sim E_B a_p$. This result turns out to give the correct parameter dependence in leading order, see eq.\ (\ref{eq:tau_crit}) below.

A more detailed stability analysis is based on the stability equations of shallow shells and allows to obtain a quantitative result including effects from a weak plate curvature. In this approach, the curved plate (or shallow shell) is described by its height profile $z(x,y) = a_c y^3 / 6$ which results in curvatures $\kappa_x = 0$ and $\kappa_y = a_c y$. The state of stress reads $\tau_y = 0$, $\tau_{xy} = 0$ and $\tau_x = -\tau_0 (1- a_p y^2)$. The numerical values of the parameters of the stress parabola and the cubic shape parabola ($\tau_0$, $ a_p$ and $ a_c$) will be calculated below from the axisymmetric buckled solution. The stability of the axisymmetric buckled solution can then be investigated by using shell stability equations \cite{Ventsel2001}, which are a set of partial differential equations for the the normal displacement $w(x,y)$ and Airy stress function $\sphi(x,y)$,
\begin{align}
 \Lap^2 \sphi &= - EH_0 \vec\nabla_\kappa^2 w \\
 E_B \Lap^2 w &= \vec\nabla_\kappa^2 \sphi + \tau_x \del_{xx}w + 
  2 \tau_{xy} \del_{xy} w + \tau_y \del_{yy} w
\end{align} 
where $\Lap = \del_{xx} + \del_{yy}$ is the Laplacian and $\vec\nabla_\kappa^2 = \kappa_y \del_{xx} + \kappa_x \del_{yy}$ is the Vlasov operator. The existence of a non-trivial solution of these stability equations indicates the existence of an unstable deformation mode for the axisymmetric buckled solution (i.e.\ a negative eigenvalue of the second variation of the elastic energy).

For the present geometry, the stability equations assume a rather simple form. In the numerical analysis, we assume wrinkles of harmonic shape in $x$-direction, $w(x,y) = W(y) \sin kx$ and $\sphi(x,y) = \Phi(y) \sin kx$. The $y$-dependent amplitude functions are to be determined, as well as the wave number $k$ of the mode which becomes unstable first. Inserting this Ansatz and the expressions for tensions and curvatures results in two coupled linear ordinary differential equations
\begin{align}
& \left(\del_y^4 - 2k^2 \del_y^2 + k^4-\frac{k^2}{E_B}\tau_0
  \left( 1- a_p y^2 \right) \right) W + \frac{ a_c y k^2}{E_B} \Phi 
  = 0 \nonumber \\
& \left( \del_y^4 - 2 k^2 \del_y^2 + k^4 \right) \Phi 
  - \left( a_c y k^2 EH_0\right) W = 0. 
\label{eq:DGL}
\end{align}
They can be solved numerically by a shooting method on the interval $y\in[0,y_\text{max})$ when we specify boundary conditions. Due to the symmetry of the problem, we expect $W(y)$ to be an even function, and from (\ref{eq:DGL}) follows directly that $\Phi(y)$ must be an odd function. Thus the starting conditions are $\Phi(0) = \Phi''(0) = 0$, $W'(0) = W'''(0) = 0$ and $W(0)=1$ (the last choice is arbitrary since the differential equations are homogeneous). We may let the plate be infinitely large, so that the wrinkles are not confined by the plate edges but by the locality of the compression. Thus, the wrinkle amplitude $W$ must approach $0$ for $y \rightarrow \infty$, as well as the tensions and, thus, the slope of the stress potential, $\Phi'(y)|_{y \rightarrow \infty}=0$. In practice, we impose $W(y_\text{max}) = W'(y_\text{max}) = 0$ and $\Phi'(y_\text{max}) = \Phi''(y_\text{max}) = 0$ for a sufficiently large $y_\text{max}$.

For the shooting method, there are only three shooting parameters among the initial conditions but four boundary conditions at the far end because the differential equations are homogeneous and, thus, the choice of $W(0)$ is arbitrary and cannot serve as a shooting parameter. Instead, we have to use one of the parameters in (\ref{eq:DGL}) as a shooting parameter. In fact, we can interpret (\ref{eq:DGL}) as an \emph{eigenvalue problem}: For given $k$, $ a_p$ and $ a_c$, find $\tau_0$ so that the differential equation has a non-trivial solution. Thus we add $\tau_0$ to the shooting parameters and have four in total, sufficient to satisfy the four boundary conditions at $y_\text{max}$.

Using this procedure, we solve (\ref{eq:DGL}) for given $E_B$, $EH_0$, $k$, $ a_p$ and $ a_c$ and determine the wrinkle amplitude $W(y)$, stress potential $\Phi(y)$ and the critical value $\tau_0$ for which a non-trivial solution exists. The wave number $k$ from our ansatz is not fixed. Since we assume the plate to be infinitely long in $x$-direction, $k$ is continuous. For our purpose, only the wrinkling mode which becomes unstable first is relevant, i.e.\ we can minimise $\tau_0$ with respect to $k$, which yields the \emph{critical tension} $\tau_c = \min_k \tau_0(k)$ and the corresponding critical wave number $k_c$ or wavelength $\lambda_c=2\pi/k_c$.

\begin{figure}
  \centerline{\includegraphics[width=85mm]{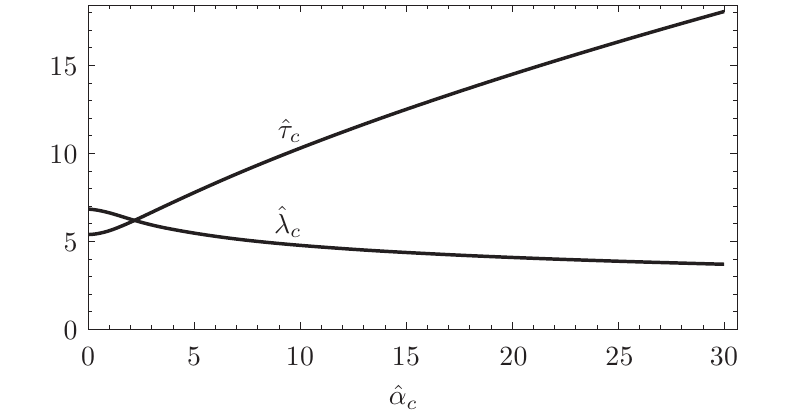}}
  \caption{Numerical results for the non-dimensional critical tension $\hat \tau_c = \tau_c/E_B a_p$ and wavelength $\hat\lambda_c = \lambda_c \sqrt{ a_p}$ as a function of the curvature parameter $\hat a_c = \sqrt{EH_0/E_B} a_c/ a_p^{3/2}$. The increase of $\hat \tau_c$ with increasing curvature parameter reflects the fact that curved plates are harder to bend in the transverse direction than flat plates (which have $\hat a_c = 0$).}
\label{fig:krumm_tau_lambda}
\end{figure}

The numerical results are plotted in fig.\ \ref{fig:krumm_tau_lambda}. Nondimensionalisation shows that the dimensionless critical stress $\hat \tau_c = \tau_c/E_B a_p$ can only depend on one other dimensionless parameter, $\hat a_c$ \cite{Knoche2014}, which describes the magnitude of the initial plate curvature. The final result of our stability analysis of a curved plate under locally compressive stress is the critical stress at which wrinkling occurs,
\begin{equation}
 \tau_c =  a_p E_B \hat\tau_c(\hat a_c) 
 \quad \text{with} \quad 
\hat a_c =\sqrt{{EH_0}/{E_B}}a_c a_p^{-3/2}
 \label{eq:tau_crit}
\end{equation} 
where the function $\hat\tau_c(\hat a_c)$ is known numerically, see fig.\ \ref{fig:krumm_tau_lambda}.

Our analysis also shows that the secondary buckling transition is a \emph{continuous} transition in the sense that the wrinkle amplitude $W$ at the transition can remain arbitrarily small \cite{Knoche2014}. This is in contrast to the primary buckling transition, which is a \emph{discontinuous} transition with metastability above and below the transition \cite{Knoche2011} and with an axisymmetric dimple of the buckled state which always has a finite size.

\section{Phase diagram for deflated spherical capsules}

The function $\hat\tau_c(\hat a_c)$ generated this way can now be applied to the stability analysis of the axisymmetric buckled capsule shapes. For a given numerical solution of the axisymmetric shape equations, we have to compute the parameters $a_p$ and $a_c$, calculate the critical buckling stress according to (\ref{eq:tau_crit}) and compare it to the minimum value $\tau_\text{min}=\min_{s_0}\tau_\phi(s_0)$ of the hoop stress in the compressive region. If $\tau_\text{min} < -\tau_c$, then the capsule cannot bear the compression and will form polygonal wrinkles, losing its axisymmetry.

The curvature parameter $a_c$ is, by definition, $a_c = \kappa_s'(s_c)$ where $s_c$ is the root of $\kappa_s$. As mentioned beforehand, the parameter $a_p$ for the parabola of the stress profile is to be determined by the condition that the approximating parabola has the same integral over the compressive region as the original stress function $\tau_\phi(s_0)$. Let $F=\int_{s_1}^{s_2} \tau_\phi(s_0) \, \diff s_0$ denote this integral, which has the physical interpretation of the net force in the compressive region $s_0 \in [s_1, s_2]$. It can be evaluated numerically for a given solution. For a parabola of the form $\tau_{\phi p}=-\tau_0\left( 1- a_p (s_0-s_c)^2 \right)$, one finds $a_p = (4\tau_0/3F)^2$, which is to be inserted into (\ref{eq:tau_crit}).

\begin{figure}
  \centerline{\includegraphics[width=85mm]{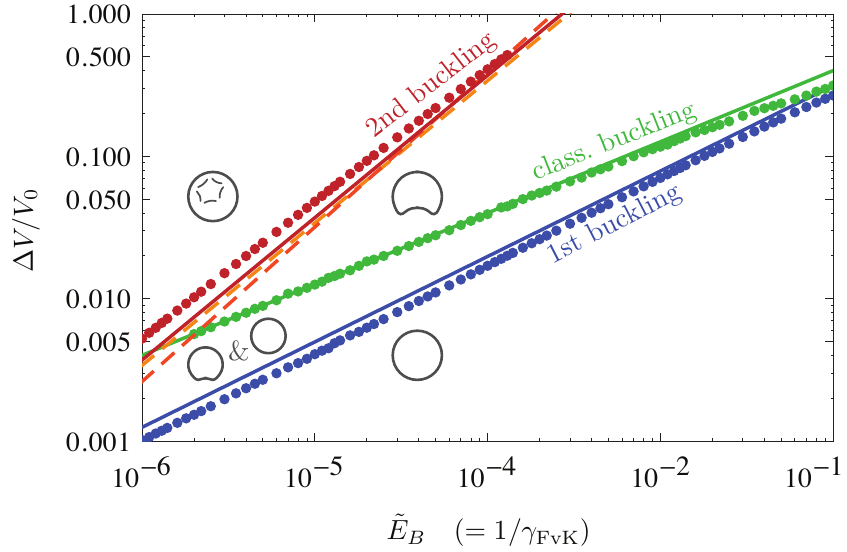}}
  \caption{(Colour online) Phase diagram of deflated spherical capsules with Poisson ratio $\nu=1/3$ in the plane of reduced bending stiffness $\tilde E_B$ and reduced volumes $\Delta V/V_0$ (double logarithmic). Dots represent results from the shape equations. The blue and red lines represent the critical volumes of first and secondary buckling, respectively, according to the Pogorelov model. The green line is the classical result for axisymmetric buckling. Dashed lines are results for the secondary buckling according to computer simulations (refs.\ \cite{Quilliet2010,Quilliet2012}).  }
\label{fig:phase_diag}
\end{figure}

In our numerical analysis, we applied this scheme to axisymmetric buckled shapes with different bending stiffnesses $\tilde E_B$ and reduced volumes $\Delta V/V_0$. We control the volume rather than the pressure, since for given pressure the capsule buckles through \cite{Knoche2011}. In this case, the secondary buckling might take place in a modified form. For each value of $\tilde E_B$, the critical capsule volume, where the criterion $\tau_\text{min} < -\tau_c$ for polygonal buckling is fulfilled, is determined numerically. This critical volume for the secondary buckling transition is shown in the phase diagram, fig.\ \ref{fig:phase_diag} (red dots). Fitting the data points with a power law (i.e.\ a straight line in the double logarithmic phase diagram) yields 
\begin{equation}
\left. ({\Delta V_\text{2nd}}/{V_0}) \right|_\text{shape eqs.} 
  = (2550 \pm 50) \, \tilde E_B^{0.946 \pm 0.002} 
\end{equation} 
with an exponent close to  $-1$.

Analysing the Pogorelov model with our secondary buckling criterion, we can also derive a simple analytical expression for the critical volume where the secondary buckling occurs. We find the following analytical results for the three parameters of the plate buckling criterion,
\begin{align*}
 \tau_0 &\sim \frac{EH_0}{1-\nu^2} \left[ \tilde E_B (1-\nu^2) \frac{\Delta V}{V_0} \right]^{1/4}, \\
a_p &\sim R_0^{-2} \left[\tilde E_B(1-\nu^2)\right]^{-1/2}, \quad
 a_c \sim \left[ \frac{1-\nu^2}{\tilde E_B} \frac{\Delta V}{V_0} \right]^{1/4}.
\end{align*}
Using these scaling results in the secondary buckling criterion (\ref{eq:tau_crit}) (treating $\hat \tau_c (\hat a_c)$ as a numerical factor) yields $\Delta V/V_0 \sim \tilde E_B (1-\nu^2)$. A detailed calculation which includes the prefactors yields 
\begin{equation}
 \left. ({\Delta V_\text{2nd}}/{V_0}) \right|_\text{Pogorelov} 
 = 3706 \, \tilde E_B,
\end{equation}
where the exponent $1$ is exact \cite{Knoche2014}; the prefactor still weakly  depends on $\nu$ and is given here for $\nu = 1/3$. This result is very close to the results from the shape equations (see fig.\ \ref{fig:phase_diag}, red line).

For both models, the number of wrinkles can be obtained by comparing the critical wavelength $\lambda_c$ (fig.\ \ref{fig:krumm_tau_lambda}) to the perimeter $2 \pi r(s_c)$ of the parallel on which the wrinkles form. In the Pogorelov model we obtain $8.6$ wrinkles, independent of the bending stiffness; using the  shape equations, we find between $8$ (for small $\tilde E_B$) and $6$ wrinkles (for larger $\tilde E_B$). A more detailed discussion is postponed to ref.\ \cite{Knoche2014}.

Figure \ref{fig:phase_diag} also shows, in dashed lines, results of computer simulations for the critical volume of the secondary buckling, which can be fitted by power laws $\Delta V/V_0 = 3400 \,\tilde E_B$ \cite{Quilliet2008,Quilliet2010} and $\Delta V/V_0 = 8470 \, \tilde E_B^{1.085}$ \cite{Quilliet2012} with exponents close to $1$. Results from our secondary buckling criterion match the simulation results fairly well but predict slightly smaller exponents $\leq 1$.

The phase diagram is supplemented by corresponding lines for the first buckling transition from a spherical to an axisymmetrically buckled shape. The classical buckling line (green line in fig.\ \ref{fig:phase_diag}) is derived from the well known classical buckling pressure $p_\text{cb}=-4 \sqrt{EH_0 E_B}/R_0^2$ \cite{Timoshenko1961,Landau1986,Ventsel2001}. It describes the pressure at which the spherical configuration becomes unstable. To convert this critical pressure into a critical volume, we have to employ the pressure-radius relation $R(p) \approx R_0 + p(1-\nu)R_0^2 / 2EH_0$ of the spherical deformation branch (valid for sufficiently small $\tilde E_B$) \cite{Knoche2011}. Hence, the volume difference at $p_\text{cb}$ is
\begin{equation}
 {\Delta V_\text{cb}}/{V_0} \approx 6 (1-\nu)  \tilde E_B^{1/2}
 \label{eq:classbuck}
\end{equation}
with an exponent $1/2$ \cite{Quilliet2012}. This line coincides very well with the data points from the shape equations (green points in fig.\ \ref{fig:phase_diag}), which were taken at the volume where the axisymmetric buckled shapes branch off the spherical shapes \cite{Knoche2011}.

The axisymmetric buckled state is unstable if pressure is controlled instead of volume \cite{Landau1986,Koiter1969,Knoche2011}, because the load that the capsule can bear is getting smaller when the dimple grows. Thus, for given pressure, the dimple that forms at the classical buckling pressure $p_\text{cb}$ grows spontaneously until a shape with stable pressure-volume relation is found. For all bending rigidities considered in ref.\ \cite{Knoche2011}, this only happens if the dimple gets in contact with the opposite side of the capsule.

Already for volume differences $\Delta V/V_0$ smaller than that of the classical buckling transition (\ref{eq:classbuck}), the spherical shape is only metastable. From the solutions of the shape equations, we can compute the smallest volume difference where the branch of axisymmetric buckled shapes becomes energetically favourable to the spherical solutions (blue points in fig.\ \ref{fig:phase_diag}) \cite{Knoche2011}. For small $\tilde E_B$, this critical volume difference is substantially smaller than that of the classical buckling transition, thus leaving a large volume region (between the two lines) where the spherical shape is metastable and the axisymmetric dimpled shape is the global energy minimum. Koiter's stability analysis \cite{Koiter1969} suggests that the buckling transition of real (imperfect) shells occurs somewhere in this region, depending on the severity of imperfections.

Pogorelov's model can also be used to calculate the volume where the elastic energies of the spherical and dimpled shapes are equal. The energy for the buckled shape is given by (\ref{eq:Upog}). For the spherical deformation, the elastic energy is, for small $\Delta V/V_0$, \cite{Quilliet2008}
\begin{equation}
 U_\text{sph} \approx \frac{4\pi}{9} \frac{EH_0 R_0^2}{1-\nu} 
  \left( \frac{\Delta V}{V_0} \right)^2. \label{eq:Usph}
\end{equation} 
Equating (\ref{eq:Upog}) to
(\ref{eq:Usph}), and solving for $\Delta V/V_0$ gives
\begin{equation}
 \left. \frac{\Delta V_\text{1st}}{V_0} \right|_\text{Pogorelov} 
= 6 J^{4/5} (1-\nu)^{4/5} (1-\nu^2)^{-1/5}  
\tilde E_B^{3/5}
\end{equation}
for the critical volume of the first buckling transition 
with an exponent $3/5$. This result is in close agreement with the data points from the shape equations (see fig.\ \ref{fig:phase_diag}, blue line).

\section{Conclusions}

In this Letter, we explained the mechanism underlying the secondary buckling instability of an intially spherical elastic capsule including a quantitatively correct value for the critical capsule volume. This completes our theoretical understanding of the generic deformation behaviour of spherical capsules upon volume reduction, which starts with a spherical shape for small volume changes, then jumps to an axisymmetric buckled shape in a primary buckling transition, and finally results in a non-axisymmetric shape with polygonal wrinkles along the inner neighbourhood of the dimple edge after the secondary buckling transition.

So far, the secondary buckling transition has only been observed in experiments or simulations but was lacking a physical explanation. The key ingredient underlying the secondary buckling is a locally compressive hoop stress, with a characteristic  negative peak near the edge of the axisymmetric dimple. We conducted a quantitative analysis, in that we approximated the profile of the compressive hoop tension $\tau_\phi$ by a parabola. This led to a derivation of a critical compressive stress, quite analogous to the critical force in the Euler buckling of bars: When the critical stress is reached, the membrane cannot support the compression any more and buckles out of its symmetric shape in order to release the compressive stress. Our analysis also showed that the secondary buckling transition is continuous as opposed to the primary buckling transition, which is discontinuous. 
This allows us to obtain a complete phase diagram (fig.\ \ref{fig:phase_diag}) which contains the stability regimes of all three relevant shapes (disregarding higher order mestastable shapes which are obtained from the shape equations \cite{Knoche2011}). 

The transition from spherical to axisymmetric buckled shape occurs at a capsule volume between the first buckling volume and the classical buckling volume. The first buckling volume is defined by the requirement that the elastic energies of the spherical and buckled shape are identical and depends on the reduced bending stiffness via $\Delta V/V_0 \sim \tilde E_B^{3/5}$. At the classical buckling volume, the spherical shape gets unstable; it reads $\Delta V/V_0 \sim \tilde E_B^{1/2}$. Between these two critical volumes, the axisymmetric buckled shape is the stable, energetically favourable state, and the spherical shape is metastable.

Applying our secondary buckling criterion to numerical axisymmetric solutions of the shape equations and to the analytic model proposed by Pogorelov, we found that the critical volume for the secondary buckling is proportional to $\Delta V/V_0 \sim \tilde E_B$. These results are in good agreement with all existing numerical simulation data except numerical results in ref.\ \cite{Vliegenthart2011}, where $\Delta V/V_0 \sim \tilde E_B^{3/4}$ is found. This differing result might be caused by using a vanishing equilibrium curvature $\kappa_{s_0} = \kappa_{\phi_0} =0$ in the elastic energy of  the simulation model in ref.\ \cite{Vliegenthart2011}.

\bibliographystyle{eplbib.bst}
\bibliography{literature_ohne_url}

\end{document}